\documentclass[seceq]{ptptex}

\usepackage{graphicx}

\makeatletter

 \@addtoreset{equation}{section}
\makeatother
\newcommand{\beq}{\begin{eqnarray}}
\newcommand{\eeq}{\end{eqnarray}}

\title{
Three dimensional conformal sigma models
}

\author{
Takeshi \textsc{Higashi}
\footnote{ E-mail: {\tt higashi@het.phys.sci.osaka-u.ac.jp}},
Kiyoshi \textsc{Higashijima}
\footnote{E-mail: {\tt higashij@phys.sci.osaka-u.ac.jp}} and  
Etsuko \textsc{Itou}
\footnote{E-mail: {\tt itou@het.phys.sci.osaka-u.ac.jp}}

}

\inst{
Department of Physics, 
Graduate School of Science, Osaka University,\\ 
Toyonaka, Osaka 560-0043, Japan
}

\abst{
We construct novel conformal sigma models in three dimensions.
Nonlinear sigma models in three dimensions are nonrenormalizable in
perturbation theory. We use the Wilsonian renormalization group equation
method to find the fixed points. The existence of fixed points is extremely
important in this approach to show the renormalizability. Conformal sigma
models are defined as the fixed point theories of the Wilsonian
renormalization group equation. The Wilsonian renormalization group
equation with anomalous dimension coincides with the modified Ricci flow
equation. The conformal sigma models are characterized by one parameter which
corresponds to the anomalous dimension of the scalar fields. Any Einstein-K\"{a}hler manifold corresponds to a conformal field theory when the anomalous dimension is $\gamma=-1/2$. Furthermore, we investigate the properties of target spaces in detail for the two dimensional case, and we find that the target space of the fixed point theory can become compact or noncompact, depending on the value of the anomalous
dimension. 
}

\begin{document}

\maketitle

\section{Introduction}
There are many perturbatively nonrenormalizable theories; gravity, gauge theories in higher dimensions and so on.
One of such theories is the three dimensional non-linear sigma model.
This model is perturbatively nonrenormalizable, but interestingly it is "renormalizable" in some nonperturbative methods.
We have investigated this model using the Wilsonian renormalization group method \cite{HI, HI3}.\\
The Wilsonian renormalization group (WRG) equation describes the infinitesimal transformation of the Wilsonian effective action when we change the ultraviolet (UV) cutoff scale $\Lambda$ to $\Lambda(\delta t)=e^{-\delta t} \Lambda$ \cite{WK, WH, Morris}.
The equation consists of two parts;
the contribution of the higher frequency modes and the terms arising from rescaling the field variables to normalize the kinetic term.
In general, the Wilsonian effective action has an infinite number of the local interaction terms, and therefore the WRG equation consists of an infinite set of differential equations for these coupling constants.
To solve the equation, we usually use the derivative expansion. The lowest order of the approximation is the local potential approximation, in which only the potential terms are retained. The next nontrivial approximation is to include the second derivative terms, generally written in the form of the sigma model Lagrangian.
In this paper, we impose ${\cal N}=2$ supersymmetry on the theory in order to forbid the appearance of the local potential terms.

In the WRG approach, the renormalizability is translated to the nontrivial continuum limit to a possible ultra-violet (UV) fixed point\cite{WK, WH, Morris}.
If there are UV fixed points, we can fine tune the coupling constant when we take the continuum limit, $\Lambda \rightarrow \infty$.
Therefore, it is important to investigate the existence of a fixed point for WRG flow.

The WRG equation for the sigma model Lagrangian gives a flow equation for the metric function of the target space. In two dimensional sigma models case, we found fixed point theories in Ref. \cite{HI2}.
One of these fixed point theories has Witten's Euclidean black hole solution as a target space \cite{Witten}.
In this paper, we will construct the three dimensional conformal sigma models using the nonperturbative renormalization group equation.
The flow equation has additional terms, in contrast to the two dimensional cases, and it corresponds to the modified Ricci flow equation \cite{Hamilton82, Chow-Knopf}.
The flow equation has one free parameter, arising from the field rescaling effects, and we show that this parameter represents the conformal dimension of the scalar field.

This paper is organized as follows\footnote{Part of this work has been reported at several conferences.\cite{HHI}}.
In section $2$ and $3$, we will shortly review of the three dimensional nonlinear sigma model, and the WRG equation for it, respectively.
In section $4$, we discuss the fixed point theory for a special value of the anomalous dimension.
We will investigate more general cases in section $5$. In this section, we will confine ourselves to two dimensional target spaces for simplicity.
Finally, we will study the conical singularity of target spaces in section $6$. In the appendix, we will discuss the fixed point theories in other coordinate system.

\section{Nonlinear sigma model with ${\cal N}=2$ supersymmetry in three dimensions}
Nonlinear sigma models with ${\cal N}=2$ supersymmetry in three dimensions are defined by the so-called K\"{a}hler potential $K(\phi, \bar{\phi})$, which is a function of the chiral and anti-chiral superfields, $\phi^i$ and $\bar{\phi}^{\bar{j}}$. A chiral superfield $\phi^i(x,\theta)$ consists of a complex scalar field $\varphi^i(x)$ and a complex fermion $\psi^i(x)$,
\[
\phi^i(x,\theta)=\varphi^i(x)+\theta\psi^i(x)+\theta^2F^i(x),
\]
where $F^i(x)$ is an auxiliary field.  The bosonic fields $\varphi^i(x)$ play the role of the coordinates of the target manifold ${\cal M}$. 
 The metric, characterizing the target manifold ${\cal M}$, is obtained as the second derivative of this K\"{a}hler potential:\cite{WB}
\[
g_{i\bar{j}}=\frac{\partial^2K(\varphi, \bar{\varphi})}{\partial \varphi^i\partial \bar{\varphi}^{\bar{j}}}\equiv K,_{i\bar{j}}.
\]
The manifold defined by a K\"{a}hler potential is called the K\"{a}hler manifold.
This metric is an arbitrary function of the scalar fields. 
The Lagrangian of the nonlinear sigma model with ${\cal N}=2$ supersymmetry reads
\begin{eqnarray}
{\cal L} = g_{i\bar{j}} \partial_{\mu} \varphi^i \partial^{\mu} \bar{\varphi}^{\bar{j}} + i g_{i\bar{j}} \bar{\psi}^{\bar{j}} (D\llap / \psi)^i + 
\frac{1}{4} R_{i\bar{j}k\bar{l}} \psi^i \psi^k \bar{\psi}^{\bar{j}} \bar{\psi}^{\bar{l}}, \label{sigma-action}
\end{eqnarray}
where the covariant derivative for the fermion fields is given by
\[
(D_{\mu}\psi)^i = \partial_{\mu} \psi^i + \partial_{\mu} \varphi^j \Gamma^i{}_{j k}\psi^k.
\]
The connection and the Riemann curvature are also written in terms of the K\"{a}hler potential $K$ as follows
\begin{eqnarray}
 {\Gamma^k}_{ij} &=& g^{k\bar{l}} g_{j\bar{l},i}
                 = g^{k\bar{l}} K,_{\,ij\bar{l}}
,\nonumber\\
 R_{i\bar{j}k\bar{l}}
 &\equiv & g_{i\bar{m}} {R^{\bar{m}}}_{\bar{j}k\bar{l}}
 =  K,_{\,i\bar{j}k\bar{l}} - g^{m\bar{n}} K,_{\,m\bar{j}\bar{l}} K,_{\,\bar{n}ik}.\nonumber
\end{eqnarray}
The Ricci curvature is defined by
\[
R_{i\bar{j}}\equiv -g^{k\bar{l}}R_{i\bar{j}k\bar{l}}.
\]

The first term in the Lagrangian (\ref{sigma-action}) contains an infinite number of derivative interactions.
All these interaction terms are perturbatively nonrenormalizable, since the scalar fields have the canonical dimension $d_{\varphi}=1/2$ in three dimensions.

\section{Renormalization group equation}
The renormalization group (RG) equation for the metric of the target manifold ${\cal M}$ in three dimensional sigma models has been derived in \cite{HI,HI3}
\begin{eqnarray}
-\frac{d}{dt} g_{i \bar{j}}
=
\frac{1}{2 \pi^2}R_{i \bar{j}} 
&+&\gamma\left( 2g_{i \bar{j}}+\varphi^k g_{i \bar{j},k} 
+\varphi^{* \bar{k}}g_{i \bar{j},\bar{k}} \right)
+\frac{1}{2}\left(\varphi^k g_{i \bar{j},k} 
+\varphi^{* \bar{k}}g_{i \bar{j},\bar{k}} \right),
\nonumber
\end{eqnarray}
where $t$ parametrizes the change of the cutoff $\Lambda \rightarrow {\rm e}^{-t}\Lambda$, and $\gamma$ denotes the anomalous dimension of the field $\phi$, introduced to normalize the field at the origin
\begin{equation}
g_{i\bar{j}}\left|_{\varphi=0}\right. =\delta_{i\bar{j}}.\label{eq:r-condition}
\end{equation}
The above RG equation, derived using the so-called K\"{a}hler normal coordinates\cite{KNC}, can be written in the covariant form 
\begin{eqnarray}
-\frac{d}{dt} g_{i \bar{j}}
=\frac{1}{2 \pi^2}R_{i \bar{j}}-g_{i\bar{j}}
+\nabla_i\xi_{\bar{j}}+\nabla_{\bar{j}}\xi_i\label{eq:ricciflow}
\end{eqnarray}
if we define the vector field 
\begin{equation}
\xi^i=\left(\frac{1}{2}+\gamma\right)\varphi^i\label{eq:vector_field}
\end{equation}
in the K\"{a}hler normal coordinates. In other coordinate systems, we have to choose a vector field corresponding to the scale transformation of the target manifold. The covariant derivative for the vector field is defined by
\begin{eqnarray}
\nabla_i\xi^k = \partial_i\xi^k + \Gamma^k{}_{ij}\xi^j,\quad
\nabla_i\xi_k = \partial_i\xi_k - \Gamma^j{}_{ik}\xi_j.\nonumber
\end{eqnarray}
The RG equation (\ref{eq:ricciflow}), called the modified Ricci flow in mathematical literature\cite{Chow-Knopf, Bakas}, describes the deformation of the target manifold of the effective theory. 

It should be emphasized that although the RG equation obtained in the
perturbation theory has the similar form with the RG equation obtained
in the Wilson's renormalization method, it is valid only in the vicinity
of the free field theory, whereas the Wilsonian RG equation can be used
to study even nontrivial conformal field theories located far from
the free field theory.

The fixed point, invariant under the change of the mass scale, is obtained by solving the equation
\begin{equation}
\frac{1}{2 \pi^2}R_{i \bar{j}}-g_{i\bar{j}}
+\nabla_i\xi_{\bar{j}}+\nabla_{\bar{j}}\xi_i=0.\label{eq:cft}
\end{equation}
The metric $g_{i\bar{j}}$ satisfying this equation defines a conformal field theory, and such a solution is called the K{\"a}hler-Ricci soliton \cite{Koiso90}.

We now introduce the parameter $c$, defined as
\beq
c\equiv \frac{1}{2}+\gamma,
\eeq
 which corresponds to the conformal dimension of the scalar fields at the fixed point.

\section{Fixed point theory for $\gamma=-\frac{1}{2}$}
When the anomalous dimension of the fields takes the specific value $-\frac{1}{2}$, the fixed point of the renormalization group equation has an extremely simple form,
\begin{equation}
\frac{1}{2 \pi^2}R_{i \bar{j}}
-g_{i\bar{j}}=0.\label{eq:special-fixed-point}
\end{equation}
Comparing this with the equation for the Einstein-K\"{a}hler manifolds,
\footnote{The parameter $\lambda$ has been introduced to satisfy the renormalization condition (\ref{eq:r-condition}) at the origin.}
\begin{equation}
R_{i \bar{j}} -h\lambda^2g_{i\bar{j}}=0,\label{eq:kahler-einstein}
\end{equation}
with a positive cosmological constant $h\lambda^2>0$, 
we find the coupling constant $\lambda$ (the inverse radius of the Einstein-K\"{a}hler manifold) of the fixed point theory is given by
\begin{equation}
\lambda^2=\frac{2 \pi^2}{h}.\label{eq:fixed-point}
\end{equation}
We thus found that any Einstein-K\"{a}hler manifold corresponds to a conformally invariant field theory, when the radius, the inverse coupling constant, takes the specific value (\ref{eq:fixed-point}).

A special class of K\"{a}hler-Einstein manifolds is provided by the
hermitian symmetric space (HSS)\cite{HN} of the form $G/H$. The compact
HSS is completely classified and listed in the following table, where
$h$ denotes the dual coxeter number of the group $G$.
\footnote{For $S^2$, $\lambda^2$ is related to the radius $a^2$ of the sphere by $\lambda^2=1/2a^2$ and $h=2$.}
\bigskip
\begin{center}
\begin{tabular}{|c|c|c|c|}
 \noalign{\hrule height0.8pt}
  $G/H$& $D=dim_{ C}(G/H)$ & $h$ \\
 \hline
 \noalign{\hrule height0.2pt}
 $SU(N)/SU(N-1)\times U(1)$&$N-1$&$N$\\
 $U(N)/U(N-M)\times U(M)$&$M(N-M)$&$N$\\
 $SO(N)/SO(N-2)\times U(1)$&$N-2$&$N-2$\\
 $Sp(N)/U(N)$ &$\frac{1}{2}N(N+1)$&$N+1$\\
 $SO(2N)/U(N)$&$\frac{1}{2}N(N-1)$&$N-1$ \\
 $E_6/SO(10)\times U(1)$&$16$&$12$\\
 $E_7/E_6 \times U(1)$&$27$&$18$\\  
 \noalign{\hrule height0.8pt}
 \end{tabular}\\
\end{center}
The metric of the HSS is explicitly constructed by using the gauge theory technique in Ref.\cite{HN}, therefore it is possible to write down the Lagrangian of conformal field theories explicitly.

\section{Two-dimensional manifold}\label{sec:teo-dim}

Although it is difficult to solve Eq.(\ref{eq:cft}) explicitly for $\gamma\ne -\frac{1}{2}$, it can be solved for the two-dimensional target space ${\cal M}$ by using a graphical method. In this section, we use real variables to describe the target manifold ${\cal M}$, and choose a special gauge in which the line element of ${\cal M}$ takes the following form\footnote{We will discuss other gauges in the appendix.}
\begin{equation}
ds^2=dr^2+e^2(r)d\phi^2.\label{eq:synchronousgauge}
\end{equation}
Since our target spaces are complex manifolds, we have assumed rotational symmetry in the $\phi$ direction corresponding to the $U(1)$ symmetry, and normalize the range of $\phi$ to $0\le \phi <2\pi$. Then $e(r)$ represents the radius of a circle for a fixed value of $r$.

Now, the components of the connection are given by
\begin{eqnarray}
&&\Gamma^r_{\phi\phi}=-e{e'},\quad \Gamma^{\phi}_{r\phi}=\frac{e'}{e},\quad 
\Gamma^{\phi}_{\phi\phi}=0\label{eq:synchronous}\\
&&\Gamma^r_{rr}=\Gamma^r_{r\phi}=\Gamma^{\phi}_{rr}=0\nonumber
\end{eqnarray}
where the prime denotes derivatives with respect to $r$.
In this coordinate system, the Ricci tensor takes the following form
\begin{eqnarray}
R_{rr}&=&R^{\phi}_{r\phi r}=-\frac{e''}{e},\quad 
R_{\phi\phi}=R^r_{\phi r\phi}=-ee'' ,\nonumber
\end{eqnarray}
Corresponding to the renormalization condition (\ref{eq:r-condition}), we impose a boundary condition for $e(r)$
\begin{equation}
{\rm lim}_{r\rightarrow 0}\frac{e(r)}{r}=1.\label{eqn:bc_origin}
\end{equation}

The fixed point of the RG equation written in terms of real coordinates corresponds to the solution of
\begin{equation}
a^2R_{ij}-g_{ij}
+\nabla_i\xi_j+\nabla_j\xi_i=0,\label{eq:cft_real}
\end{equation}
where 
\[
a^2=\frac{1}{2\pi^2}.\label{eq:radius_sphere}
\]
Now, we have to find the vector field $\xi^i=(\xi^r, \xi^{\phi})$. 
This vector field $\xi^i$, representing an infinitesimal scale transformation of the target space, has to be proportional to $(cr,0)$ at least around the origin $r=0$, where the renormalization condition (\ref{eq:r-condition}) is imposed. Since we assume the rotational symmetry ($U(1)$), it is natural to assume $\xi_{\phi}=0$. Then the vector field in this coordinate system is fixed by the consistency of the coupled differential equation (\ref{eq:cft_real}).  The equation for $r-r$ and $\phi-\phi$ components are
\begin{eqnarray}
&&-a^2\frac{e''}{e}-1+2\xi'_r=0,\label{eqn:rr}\\
&&-a^2ee''-e^2+2e'e\xi_r=0.\label{eqn:phiphi}
\end{eqnarray}
The equation for the $r-\phi$ component is satisfied trivially. Then, we find 
\begin{equation}
\frac{\xi'_r}{\xi_r}=\frac{e'}{e}\label{eqn:consistency_rg}
\end{equation}
by requiring the compatibility of the two equations. 
Since $\xi_r$ has to be proportional to $cr$ near the origin, we choose the vector field $\xi_i$ in this coordinate system as
\begin{equation}
\xi^r=ce(r),\quad \xi^{\phi}=0, \qquad (c=\frac{1}{2}+\gamma),
\end{equation}
taking into account the boundary condition (\ref{eqn:bc_origin}).
Finally, we obtain the RG equation in this gauge
\begin{equation}
-a^2e''-e + 2cee'=0.\label{eq:rg_synchronous}
\end{equation}

When $c=0$, namely for $\gamma=-1/2$, the solution of this equation is easily obtained
\[
e(r)=a\sin{\frac{r}{a}},
\]
which defines the line element of the sphere $S^2$ with radius $a$,
\begin{equation}
ds^2=dr^2+a^2\sin^2{\frac{r}{a}}d\phi^2,\nonumber
\end{equation}
in conformity with the result of the previous section. 

On the other hand, when $c\ne 0$, it is convenient to rewrite the above second order 
differential equation as a set of first-order differential equations,
\begin{eqnarray}
e'&=&p \label{eq:1st_order}\\
p'&=&-\frac{1}{a^2}e(1-2cp)\nonumber
\end{eqnarray}
with the boundary conditions
\begin{equation}
e(0)=0,\quad p(0)=1.\label{eq:bc_sg}
\end{equation}

Furthermore, if we introduce the new variable
\begin{equation}
Q(r)=\frac{1}{2c}\log{|1-2cp(r)|}\ \ \mbox{and}\ \ \quad P(r)=e(r),\label{eq:def_QP}
\end{equation}
the first-order differential equations (\ref{eq:1st_order}) can be rewritten in the form of Hamilton's equation of motion:
\begin{eqnarray}
\frac{dQ}{dr}&=&\frac{P}{a^2}=\frac{\partial H}{\partial P},\nonumber
\\
\frac{dP}{dr}&=&\frac{1-{\rm e}^{2cQ}}{2c}=-\frac{\partial H}{\partial Q}.
\label{eq:canonical}
\end{eqnarray}
Here, the "Hamiltonian" is given by
\begin{equation}
H(Q,P)=\frac{1}{2a^2}P^2+V(Q),
\label{eq:hamiltonian}
\end{equation}
where $V(Q)$ represents the "potential energy"
\begin{equation}
V(Q)=-\frac{1}{2c}Q+\frac{1}{(2c)^2}{\rm e}^{2cQ}.\label{eq:potential}
\end{equation}
In Eqs.(\ref{eq:canonical}) and (\ref{eq:potential}), we have assumed $1-2cp(r)>0$. When $1-2cp(r)<0$, we have to introduce an extra minus sign in front of ${\rm e}^{2cQ}$ in these equations.

Once we fix the value of the parameter $c$, the sign of $1-2cp(r)$ at
$r=0$ is determined by the initial condition (\ref{eq:bc_sg}).
It is easily seen that $1-2cp(r)$ does not change sign at any "time "
$r$. To see this, let us assume $c>0$. (A similar argument holds for $c<0$.)
Suppose $1-2cp(r)$ changes sign at some "time" $r$. Then $Q$ has to go
to $-\infty$ at that "time" $r$, by the definition of $Q$ given in (\ref{eq:def_QP}).
However, as we will discuss below, $Q$ has a lower bound, and therefore
the sign of $1-2cp(r)$ does not change at any "time" $r$.

\begin{figure}
\begin{center}
\resizebox{!}{60mm}{\includegraphics{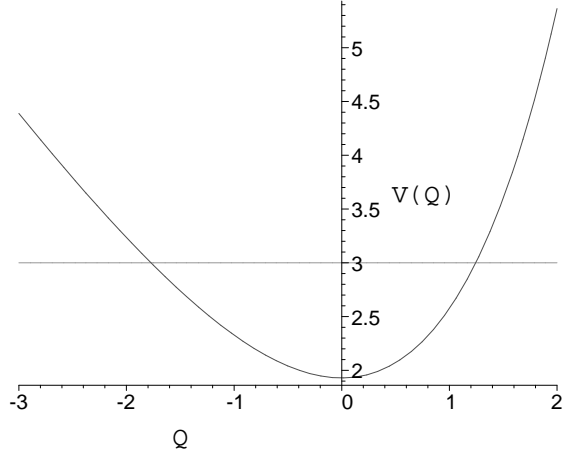}}
\end{center}
\caption{The "potential energy" $V(Q)$ in Eq.(\ref{eq:potential}) for $1-2cp(r)>0$. 
}
\label{fig:poten_sync}
\end{figure}

Since this "Hamiltonian" does not depend on $r$ explicitly, the energy is independent of the "time" $r$. We plot the potential $V(Q)$ in the case $1-2cp(r)>0$ in Fig.\ref{fig:poten_sync}. The horizontal line represents a constant value of the energy. Since the kinetic energy is positive definite, solutions exist only in the bounded region $Q_{min}\le Q \le Q_{max}$, where $Q_{min}$ and $Q_{max}$ are determined by $V(Q)=E$. It is also true for $1-2cp(r)<0$ that $Q$ has a lower bound, even when the sign of ${\rm e}^{2cQ}$ has been changed in the potential (\ref{eq:potential}). According to the initial condition (\ref{eq:bc_sg}), $p(r)$ starts from $p(0)=1$. The fact that $1-2cp(r)<0$ does not change sign implies that $p(r)<1$ if $2c<1$, and $p(r)>1$ if $2c>1$.

\begin{figure}[h]
\begin{center}
\unitlength=1mm
\begin{picture}(80,60)
\includegraphics[width=8cm]{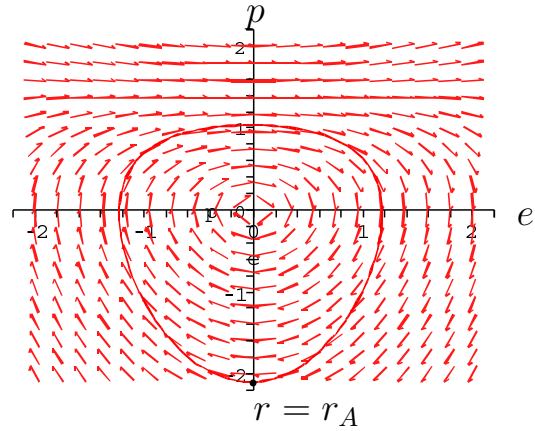}
\put(-41,55){\Large $p$}
\put(-5,28){\Large $e$}
\put(-40,2){\Large $r=r_A$}
\put(-41.3,4.8){\Huge $\cdot$}
\end{picture}
\caption{Flow of the first-order differential equations (\ref{eq:1st_order}) for $0< 2c<1$ in the "phase space" $(e(r),p(r))$. 
The solid curve represents the solution specified by the boundary condition.}\label{fig:dsphere}

\end{center}
\end{figure}

The boundary condition (\ref{eq:bc_sg}) gives the initial condition
\begin{equation}
Q(0)=\frac{1}{2c}\log{|1-2c|},\quad P(0)=0,\label{eq:initial_value}
\end{equation}
which determines the value of the energy
\begin{equation}
H(Q,P)=E=\frac{1}{(2c)^2}\left(|1-2c|-\log{|1-2c|}\right).\label{eq:energy}
\end{equation}
Then, the conservation of "energy" is expressed as follows
\begin{equation}
\frac{2}{(2c)^2}\left(-2c(1-p(r))+\log{|1-2cp(r)|}-\log{|1-2c|}\right)=\frac{e^2(r)}{a^2},\label{eq:energy_cons}
\end{equation}
which determines the closed contour shown in Fig. \ref{fig:dsphere} for $2c<1$.

\begin{figure}[h]
\begin{center}
\unitlength=1mm
\begin{picture}(80,60)
\includegraphics[width=8cm]{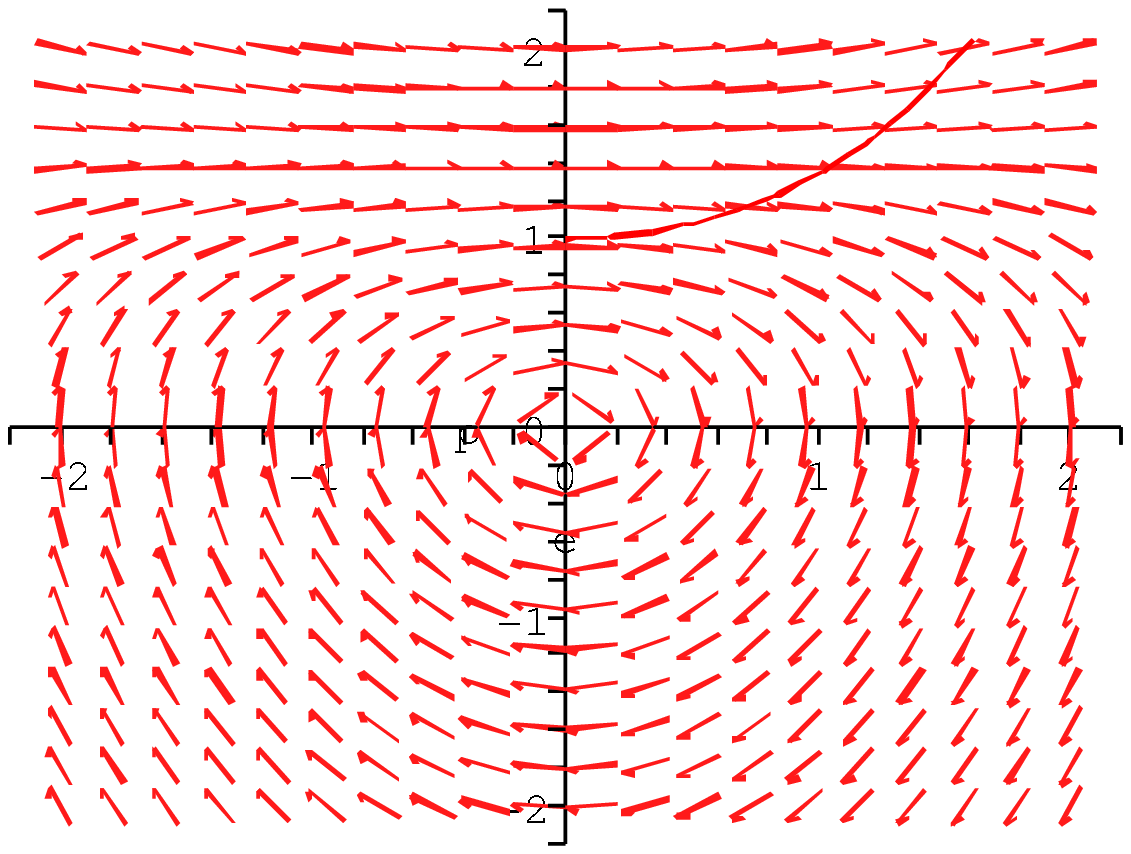}
\put(-41,55){\Large $p$}
\put(-5,28){\Large $e$}
\end{picture}
\caption{Flow of the first-order differential equations (\ref{eq:1st_order}) for $2c\geq1$ in the phase space. 
The solid curve represents the solution specified by the boundary condition.}\label{fig:non-cpt}

\end{center}
\end{figure}

The vector field of the flow (\ref{eq:1st_order}) is shown in Fig.\ref{fig:dsphere}. When $0\le 2c < 1$, this equation defines a compact manifold, since the trajectory starting from the initial point (\ref{eq:bc_sg}) comes back to $e=0$ at a finite $r=r_A$, implying that the circumference of the circle at that $r$ vanishes. 
We call this the "deformed shpere". Its metric function is approximated by 
\begin{equation}
e(r)= a \sin \left( \frac{r}{a} \right)+\frac{ca}{3} \left( 2 \sin \left( \frac{r}{a} \right) - \sin \left( \frac{2r}{a} \right) \right)+O(c^2)\label{eq:smallc}
\end{equation}
for small $c$. Similarly, we have a compact target space for $c<0$.

On the other hand, Fig.\ref{fig:non-cpt} shows the radius $e(r)$ becomes larger and larger when $r$ goes to infinity. Thus the solution corresponds to a noncompact manifold 
for $2c\ge 1$. 
To see the asymptotic behavior of $e(r)$ for large $r$, we can ignore the second term in Eq.(\ref{eq:rg_synchronous}). We then obtain 
\[
\frac{de}{dr}=1+\frac{c}{a^2}e^2,
\]
which can be integrated to obtain $e(r)$: 
\[
e(r)=\frac{a}{\sqrt{c}}\tan{\left(\frac{\sqrt{c}}{a}r\right)}.
\]
Since $e(r)$ defines the radius of the circle when the geodesic distance from the origin $r$ is fixed, this asymptotic behavior implies that for $2c>1$ the radius of the manifold increases with $r$.

\section{The Euler number and the volume of the manifold}
Let us discuss the change of the volume of the target manifold along the flow of the RG equation.
The change of the volume of the manifold along the renormalization group flow (\ref{eq:ricciflow}) is given by
\begin{eqnarray}
&&\frac{d}{dt}\int \sqrt{\det{(g)}}d\varphi d\varphi^* =\frac{1}{2}
\int\sqrt{\det{(g)}}{\rm Tr}\left(g^{i\bar{j}}\frac{dg_{i\bar{j}}}{dt}\right)
d\varphi d\varphi^*\nonumber\\
&&=-\frac{1}{4\pi^2}\int \sqrt{\det{(g)}}Rd\varphi d\varphi^*
+\frac{dim_{ C}{\cal M}}{2} \int \sqrt{\det{(g)}}d\varphi d\varphi^*,\label{volume-flow}
\end{eqnarray}
where use has been made of the covariant constancy of the metric, $\nabla g=0$, and the scalar curvature  $R$ is defined by
\[
R=g^{i\bar{j}}R_{i \bar{j}}.
\]

At the fixed point, the left hand side of the equation (\ref{volume-flow}) vanishes, and we obtain the following relation:
\beq
\frac{1}{4\pi^2}\int \sqrt{\det{(g)}}Rd\varphi d\varphi^*
=\frac{dim_{ C}{\cal M}}{2} \int \sqrt{\det{(g)}}d\varphi d\varphi^*. \label{eq:relation}
\eeq
In two dimensions, the left hand side corresponds to the Euler number when the manifold has neither a boundary nor a singularity:
\[
\chi({\cal M})=\frac{1}{4\pi}\int\sqrt{\det{(g)}}Rd\varphi d\varphi^*
=2(1-g),
\]
where the genus is $0$ for $S^2$ and $1$ for $T^2$.
Thus, the volume of the target manifold is defined by the topological quantity $\chi({\cal M})$ at the fixed point:
\[
V({\cal M})=\int\sqrt{\det{(g)}}d\varphi d\varphi^*
=\frac{2}{\pi}\chi({\cal M}).
\]
Actually, when the manifold is a round sphere $S^2$ with radius $a$,
\[
V(S^2)=4\pi a^2=\frac{2}{\pi}
\]
implies that the radius at the fixed point is 
\beq
a^2=\frac{1}{2\pi^2}.\label{S2-radius}
\eeq
This radius coincides with that appearing in the second footnote of \S. $4$.

We next consider cases in which $c \ne 0$.
Unfortunately, in such cases, the target space has a conical singularity.
Then the relationship between the Euler number and the volume is no longer valid. We show this point in detail.

In the previous section, Fig. \ref{fig:dsphere} shows that the target space with nonzero $c$ describes a deformation from $S^2$.
Then the Euler number should not change.
To show the invariance of the Euler number under this deformation, we briefly review of the Gauss-Bonnet theorem.

\begin{figure}[h]
\begin{center}
\unitlength=1mm
\begin{picture}(110,50)
\includegraphics[width=10cm]{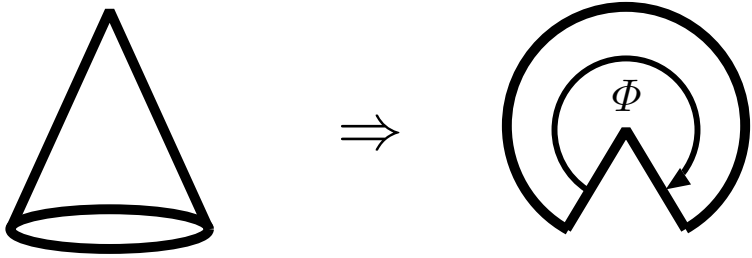}
\put(-18.5,20){\LARGE $\Phi$}
\put(-55,15){\Huge $\Rightarrow$}
\end{picture}
\caption{The deficit angle ($2 \pi-\Phi$) of the cone.}\label{fig:cone}

\end{center}
\end{figure}

Let us consider the metric of the target space with a conical singularity.
The flat space has the following metric:
\beq
ds^2=dr^2+r^2 d \phi^2.
\eeq
The cone (Fig. \ref{fig:cone}) has the additional condition 
\beq
0 \leq \phi \leq \Phi ,
\eeq
because of the deficit angle.
We rescale the angular coordinate as $\phi \rightarrow \phi'=\frac{2 \pi}{\Phi} \phi$. Then the effect of the deficit angle appears in the metric as follow:
\beq
ds^2= dr^2+\Big( r^2 \big(\frac{\Phi}{2\pi}\big)^2 \Big) d{\phi'}^2.
\eeq
In Eq.(\ref{eq:synchronousgauge}), we have set $ds^2=dr^2 +e^2(r)d\phi^2$ and used the initial condition $e'(r=0)=1$.
If the manifold is smooth at the point ($r=r_{A}$) in Fig.\ref{fig:dsphere}, the first differential of the function $e(r)$ should equal to $ -1$.
In other words, if the target space has a conical singularity at $r=r_A$, we can obtain the equation
\beq
\frac{\partial e(r)}{\partial r}|_{r=r_A}=  -\frac{\Phi}{2 \pi} \ne -1.
\eeq

In general, if the target space has several conical singularities at the points $r=r_i$, with $i=1,\cdots, n$,then, the Gauge-Bonnet theorem is given as follows:
\beq
2 \pi \chi({\cal M})=\frac{1}{2} \int_{\cal M} d V \sqrt{g} R +\sum_{i=1}^{n} (2\pi-\Phi_i).\label{Gauss-Bonnet}
\eeq
In this coordinate system, the Ricci scalar and the determinant of the metric are given by $R=-2 e''(r)/e(r)$ and $\sqrt{g}=e(r)$ respectively.
Then the first term of Eq.(\ref{Gauss-Bonnet}) is 
\beq
\frac{1}{2} \int_{\cal M} d V \sqrt{g} R= -2\pi [e'(r)|_{r=r_A}-1].
\eeq
The deficit angle at $r=r_A$ is given by
\beq
2\pi-\Phi= 2 \pi-\left( -2\pi e'(r)|_{r=r_A} \right).
\eeq
Then, we see that the Euler number remains unchanged by the $c$-deformation:
\beq
\chi({\cal M})=2.
\eeq
However, the volume of the target space depends on the value of $e'(r)$ at the point $A$. From Eq.(\ref{eq:relation}), this volume is given by
\beq
V({\cal M})=\frac{2}{\pi}\left[1-e'(r)|_{r=r_{A}} \right],
\eeq
when $0 \leq c<1/2$.
In this region, $e'(r)|_{r=r_A}$ is negative, and the absolute value monotonically increases with $c$.
For small $c$, the lowest order approximation, (\ref{eq:smallc}), gives $e'(a\pi)=-(1+\frac{4c}{3})$, which implies $\Phi=2\pi(1+\frac{4c}{3})$ for small $c$.
At $c=1/2$, the target space becomes flat space, and then the volume goes to infinity.
\section{Summary}
Three dimensional non-linear sigma models are perturbatively nonrenormalizable.
Also, we derived the non-perturbative renormalization group equation for these models in Ref.\cite{HI, HI3}. In this paper, we rewrote the renormalization group equation in covariant form using the K{\"a}hler normal coordinates.
The covariant renormalization group equation has the same form as the modified Ricci flow.
We have also investigated the fixed point theories for the renormalization group equation.

The covariant RG equation has one free parameter ($c$). The parameter corresponds to the conformal dimension of the scalar fields, and is the sum of the canonical dimension ($1/2$) and the anomalous dimension\footnote{Anomalous dimensions for some theories may be negative. Negative anomalous dimensions are allowed in nonlinear sigma models\cite{HI4}. }.
We find that the target spaces of the fixed point theories must satisfy the Einstein-K{\"a}hler condition with the specific value of the radius $c=0$.
We will discuss the stability of these fixed point target spaces in Ref.\cite{HHI2}.

For $0<c<1/2$, the target space of the fixed point theory becomes a "deformed sphere". We verified this numerically in the case of one complex dimension.
On the other hand, for $c\ge1/2$, we find that the target space is non-compact.
At the critical value $c=1/2$, the target space reduces to a flat space corresponding to a free field theory.

\section*{Acknowledgements}
We would like to thank Muneto Nitta, Ioannis Bakas and Toshiki Mabuchi
for illuminating discussions. 
This work was supported in part by Grants-in-Aid for Scientific Research (\#16340075 and \#13135215). 

\begin{appendix}
\section{Ricci Flow Equation in Other Gauges} 
%
In this appendix, we study the Ricci flow equation in two dimensions using other coordinate system, parametrized by

\begin{equation}
ds^2=A(x)(dx^1)^2+B(x)(dx^2)^2,\label{eq:def_2dmetric}
\end{equation}
which implies
\begin{equation}
g_{11}=A(x),\quad g_{22}=B(x),\quad g^{11}=\frac{1}{A},\quad g^{22}=\frac{1}{B}.
\label{eq:2dmetric}
\end{equation}

Using the definition of connections
\begin{equation}
\Gamma^k_{ij}=\frac{1}{2}g^{mk}\{\partial_jg_{im}
+\partial_ig_{jm}-\partial_mg_{ij}\},\label{eq:def_connection}
\end{equation}
various components of connections are calculated as

\begin{eqnarray}
&&\Gamma^1_{11}=\frac{\dot{A}}{2A},\quad \Gamma^1_{12}=\frac{A'}{2A},\quad  
\Gamma^1_{22}=-\frac{\dot{B}}{2A},\label{eq:connection} \\
&&\Gamma^2_{11}=-\frac{A'}{2B},\quad \Gamma^2_{12}=\frac{\dot{B}}{2B},\quad  
\Gamma^2_{22}=\frac{B'}{2B},\nonumber
\end{eqnarray}
where
\[
\dot{A}\equiv \frac{\partial A}{\partial x^1},\quad 
A'\equiv \frac{\partial A}{\partial x^2}.
\]

Various Riemann curvatures are given by
\begin{eqnarray}
R^1_{212}&=&-\partial_1\left(\frac{\dot{B}}{2A}\right)-\frac{1}{2}(\log{A})''
-\frac{A^{'2}}{4A^2}+\frac{{\dot{B}}^2}{4AB}-\frac{\dot{A}\dot{B}}{4A^2}
+\frac{A'B'}{4AB},\label{eq:rieman}\\
R^2_{121}&=&-\frac{1}{2}\partial^2_1\log{B}-\partial_2\frac{A'}{2B}
+\frac{A^{'2}}{4AB}-\frac{{\dot{B}}^2}{4B^2}+\frac{\dot{A}\dot{B}}{4AB}
-\frac{A'B'}{4B^2},\nonumber
\end{eqnarray}
where we have used the definition
\begin{equation}
R^{\ell}_{ijk}=\partial_j\Gamma^{\ell}_{ik}-\partial_k\Gamma^{\ell}_{ij}
+\Gamma^m_{ik}\Gamma^{\ell}_{mj}-\Gamma^m_{ij}\Gamma^{\ell}_{mk}.
\label{eq:def_riemann}
\end{equation}
Ricci and scalar curvatures are obtained using the Riemann curvature.

\noindent
{\large\bf Ricci tensor:}
\begin{eqnarray}
R_{11}=R^2_{121},\quad R_{22}=R^1_{212},\quad R_{12}=R_{21}=0.
\end{eqnarray}

\noindent
{\large\bf Scalar curvature:}
\begin{equation}
R=g^{ij}R_{ij}=\frac{R_{11}}{A}+\frac{R_{22}}{B}.\label{eq:scalar_curvature}
\end{equation}

\subsection{Robertson-Walker type gauge}
If the system has rotational symmetry, we can safely assume the following form of the metric:
\[
x^1=r,\quad x^2=\phi,\quad A(x)=f^2(r),\quad B(x)=r^2,
\]
then, the line element is written as follows
\begin{equation}
ds^2=f^2(r)dr^2+r^2d\phi^2,\label{eq:def_2dmetric_robertson}
\end{equation}
where $r$ denotes the radius of the circle when $r$ is fixed.
\[
g_{rr}=f^2,\quad g_{\phi\phi}=r^2,\quad g^{rr}=\frac{1}{f^2},\quad g^{\phi\phi}=\frac{1}{r^2}.
\]
The connections and curvatures in this gauge are given by
\begin{eqnarray}
&&\Gamma^r_{rr}=\frac{f'}{f},\quad \Gamma^{\phi}_{r\phi}=\frac{1}{r},\quad  
\Gamma^r_{\phi\phi}=-\frac{r}{f^2},\quad\mbox{others}=0,\label{eq:connection}\\ 
&&R_{rr}=R^{\phi}_{r\phi r}=\frac{f'}{rf},\quad 
R_{\phi\phi}=R^r_{\phi r\phi}=\frac{rf'}{f^3},
\end{eqnarray}
where the prime denotes the derivative with respect to $r$.

The vector field of the rescaling is given in this coordinate system by
\begin{equation}
\xi^r=\frac{cr}{f(r)},\quad \xi^{\phi}=0, \nonumber
\end{equation}
where
\[
c=\frac{1}{2}+\gamma.
\]
Since the covariant components are
\[
\xi_r=crf,\quad \xi_{\phi}=0,
\]
the covariant derivatives of the vector field,
\begin{eqnarray}
\nabla_r\xi_r&=&\frac{\partial}{\partial r}(crf)-\Gamma^r_{rr}(crf)=cf,\nonumber\\
\nabla_{\phi}\xi_{\phi}&=&-\Gamma^r_{\phi\phi}(crf)=c\frac{r^2}{f},\nonumber
\end{eqnarray}
lead to an identical differential equation for the $rr$- and $\phi\phi$- 
components of the RG equation
\begin{equation}
a^2\frac{f'}{rf}-f^2+2cf=0, \quad\mbox{where}\quad a^2=\frac{1}{2\pi^2}.\label{eq:rg_rwmetric}
\end{equation}
By a change of the variable $x=r^2$, it can be rewritten as
\begin{equation}
2a^2\frac{df}{dx}-(f-2c)f^2=0,\label{eq:rgx_rwmetric}
\end{equation}
which can be integrated to give
\[
\frac{2}{(2c)^2}\left(\frac{2c}{f}
+\log{|\frac{f-2c}{f}|}\right)=\frac{x}{a^2}+const.
\]
The boundary condition $f(0)=1$ fixes the integtation constant:
\begin{eqnarray}
\frac{2}{(2c)^2}\left(\frac{2c}{f}-2c
+\log{|\frac{f-2c}{f}|}-\log{|1-2c|}\right)=\frac{x}{a^2}.
\label{eq:implicit}
\end{eqnarray}
When $c=0$ ($\gamma=-\frac{1}{2}$ case), the solution of this implicit equation is very simple,
\begin{equation}
f(r)=\frac{1}{\sqrt{1-\frac{r^2}{a^2}}}.\nonumber
\end{equation}
This coordinate covers half of the round sphere $S^2$. To cover the whole 
sphere, we introduce the new coordinate $\theta$ with the range $0\le \theta \le \pi$:
\[
ds=a^2(d\theta^2+\sin^2{\theta}d\phi^2),\quad r=a\sin\theta.
\]

\begin{figure}
\begin{center}
\resizebox{!}{60mm}{\includegraphics{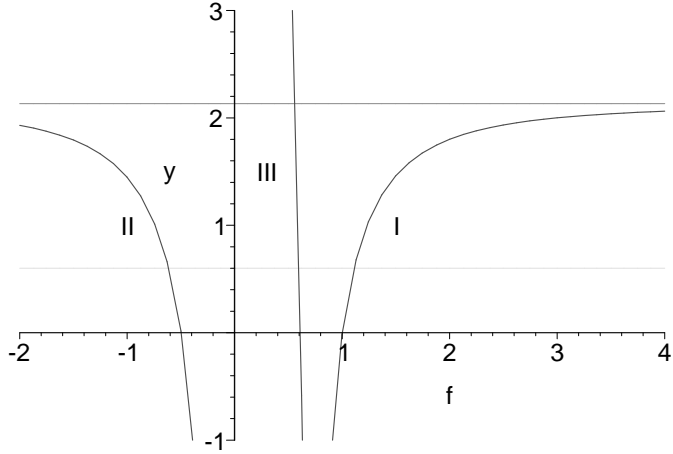}}
\end{center}
\caption{The left-hand side of Eq.(\ref{eq:implicit}) for $c=0.36$.}
\label{fig:lefthand}
\end{figure}

For a non-vanishing value of $c$, namely when $\gamma\ne -\frac{1}{2}$, 
the left hand-side of the implicit equation (\ref{eq:implicit}) is plotted in the Fig. \ref{fig:lefthand}. The right-hand side is represented by a horizontal line. 
For a fixed value of $x$, $f(x)$ is given by the intersection of the graph and the horizontal line $y=\frac{x}{a^2}$. The boundary condition forces us to choose the branch {\rm\bf I} near $x\approx 0$ (the south-pole). The function $f(x)$ is increasing for small $x$.
There is a maximum value of $x$ where $f(x)$ diverges to infinity. As in the case of the sphere for $c=0$, the geodesic distance to this point is finite, and we can extend further by using the solution for negative values of $f$ in the branch {\rm\bf II}. The north pole corresponds to a point where the graph of the branch {\rm\bf II} interesects with the horizontal axis $(x=0)$.

The approximate solution in the branch {\bf\rm I} of (\ref{eq:implicit}) near the origin $x=0$ is given by the Taylor series 
\begin{equation}
f(x)=1+\frac{1}{2}\cdot\frac{1-2c}{a^2}x+\frac{1}{8}\cdot\frac{3-10c+8c^2}{a^4}x^2+\cdots .\label{eq:seriessol}
\end{equation}

The maximum value of $x$ is given by
\begin{equation}
x_{max}=a^2\left(-\frac{1}{c}-\frac{1}{2c^2}\log{|1-2c|}\right)
\end{equation}
When $x$ approaches to $x_{max}$, $f(x)$ diverges as
\begin{equation}
f(x)\approx \frac{a^2}{\sqrt{x_{max}-x}}.\label{eq:asymptotic}
\end{equation}
The line element near $r\approx r_{max}=\sqrt{x_{max}}$ is
\begin{equation}
ds^2=\frac{a^2}{r_{max}^2-r^2}(dr)^2+r^2(d\phi)^2,\label{asym_line}
\end{equation}
which can be rewritten as
\begin{equation}
ds^2=a^2(d\theta)^2+r_{max}^2\sin^2{\theta}(d\phi)^2\label{eq:asym_line2}
\end{equation}
if we define the angle $\theta$ by $r=r_{max}\sin{\theta}$.
The point $r=r_{max}$, corresponding to $\theta=\pi/2$, is not a singularity but is located a finite distance from the origin $(r=0)$, and thus we can extend further to the region $\pi/2\le \theta$, where $r$ begins to decrease again.

\begin{figure}
\begin{center}
\resizebox{!}{80mm}{\includegraphics{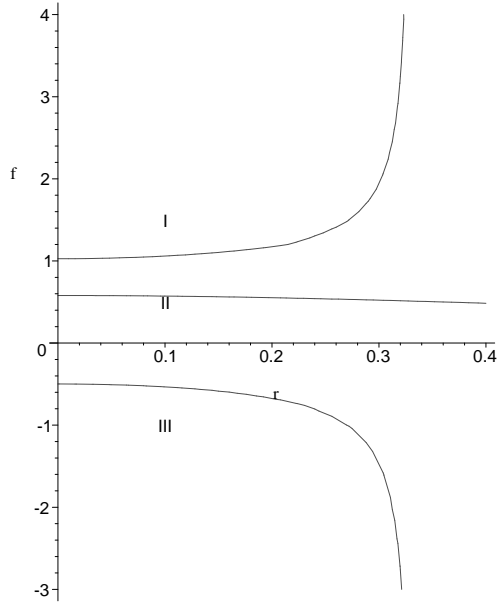}}
\end{center}
\caption{Robertson-Walker type metric $f(r)$ for $c=0.36$.}
\end{figure}

\subsection{Relations between various gauges}
Let us first discuss the relation between the Robertson-Walker type gauge (\ref{eq:def_2dmetric_robertson}) and the synchronous gauge (\ref{eq:synchronousgauge}), discussed in section \ref{sec:teo-dim}.
In this subsection, we denote the radius in the synchronous gauge as $R$.
If we define $R$ in the Robertson-Walker gauge by
\begin{equation}
R(r)=\int_0^rf(r)dr,\label{eq:def_R}
\end{equation}
the line element takes the form
\begin{eqnarray}
ds^2&=&f^2(r)dr^2+r^2d\phi^2,\nonumber\\
&=&dR^2+e^2(R)d\phi^2,\nonumber
\end{eqnarray}
where $e(R)$ defined by
\begin{equation}
e(R)=r(R),\label{eq:def_e}
\end{equation}
is the inverse of Eq.(\ref{eq:def_R}).
From Eqs.(\ref{eq:def_e}) and (\ref{eq:def_R}), we have the relation
\begin{equation}
p(R)\equiv e'(R)=\frac{dr}{dR}=\frac{1}{\frac{dR}{dr}}=\frac{1}{f(r)}.
\label{eq:def_p}
\end{equation}
If we substitute Eqs.(\ref{eq:def_e}) and (\ref{eq:def_p}) into Eq.(\ref{eq:energy_cons}), we recognize that Eq.(\ref{eq:energy_cons}) coincides with Eq.(\ref{eq:implicit}) exactly.

\bigskip
Let us next discuss the relation between the Robertson-Walker type gauge and the conformal gauge in the case that there is rotational symmetry. 
If $R$ and $\rho(R)$ satisfy 
\begin{eqnarray}
f(r)\frac{dr}{dR}=\rho(R),\quad r=R\rho(R),\label{eq:rwc_relation}
\end{eqnarray}
the line element (\ref{eq:def_2dmetric_robertson}) can be written in the form of the conformal gauge (\ref{eq:conformal_gauge}),
\begin{equation}
ds^2=\rho^2(R)(dR^2+R^2d\phi^2).\label{eq:conformal_polar}
\end{equation}

From (\ref{eq:rwc_relation}), we obtain
\begin{eqnarray}
R(r)&=&r{\rm e}^{\int_0^r\frac{f(r)-1}{r}dr},\label{eq:rw_c}\label{eq:Rr_c}\\
\rho(R)&=&\frac{r(R)}{R},\label{eq:rhoR}
\end{eqnarray}
where $r(R)$ is the inverse of Eq.(\ref{eq:Rr_c}).

In the conformal gauge,
\[
A(x)=B(x)=\rho^2(x),
\]
the line element takes the very simple form
\begin{equation}
ds^2=\rho^2(x)(dx_1^2+dx_2^2).\label{eq:conformal_gauge}
\end{equation}
The connections and curvatures in this gauge are given by
\begin{eqnarray}
&&\Gamma^1_{11}=-\Gamma^1_{22}=\Gamma^2_{12}=\frac{\dot{\rho}}{\rho},\label{eq:conformal}\\
&& \Gamma^1_{12}=-\Gamma^2_{11}=\Gamma^2_{22}=\frac{{\rho}'}{\rho},\\
&& R^1_{212}=R^2_{121}=-\Delta \log{\rho},\\
\label{eq:riemann_curvature_conformal}
&& R_{11}=R_{22}=-\Delta \log{\rho},\\
\label{eq:ricci_conformal}
&& R=R_{11}+R_{22}=-\frac{2}{\rho^2}\Delta \log{\rho}.
\label{eq:scalar_curvature_conformal}
\end{eqnarray}
\end{appendix}

\end{document}